\documentclass[prl,twocolumn,floatfix]{revtex4}
\pdfoutput=1

\usepackage{graphicx}
\usepackage{color}
\usepackage{ulem}

\begin{document}
\title{Quantum-inspired interferometry with chirped laser pulses}
\author{R. Kaltenbaek}
\email{rkaltenb@iqc.ca}
\affiliation{Institute for Quantum Computing and Department of Physics \& 
Astronomy, University of Waterloo, Waterloo, Canada, N2L 3G1}
\author{J. Lavoie}
\affiliation{Institute for Quantum Computing and Department of Physics \& 
Astronomy, University of Waterloo, Waterloo, Canada, N2L 3G1}
\author{D.N. Biggerstaff}
\affiliation{Institute for Quantum Computing and Department of Physics \& 
Astronomy, University of Waterloo, Waterloo, Canada, N2L 3G1}
\author{K.J. Resch}
\email{kresch@iqc.ca}
\affiliation{Institute for Quantum Computing and Department of Physics \& 
Astronomy, University of Waterloo, Waterloo, Canada, N2L 3G1}

\begin{abstract}
We introduce and implement an interferometric technique based on
chirped femtosecond laser pulses and nonlinear optics.  The
interference manifests as a high-visibility ($>85$\%)
phase-insensitive dip in the intensity of an optical beam when the
two interferometer arms are equal to within the coherence length of
the light. This signature is unique in classical interferometry, but
is a direct analogue to Hong-Ou-Mandel quantum interference. Our
technique exhibits all the metrological advantages of the quantum
interferometer, but with signals at least $10^7$ times greater. In
particular we demonstrate enhanced resolution, robustness against
loss, and automatic dispersion cancellation. Our interferometer
offers significant advantages over previous technologies, both
quantum and classical, in precision time delay measurements and
biomedical imaging.
\end{abstract}

\pagestyle{empty}
\maketitle

Interference is a defining feature
of both quantum and classical theories of light. It also enables the
most precise measurements of a wide range of physical quantities
including length~\cite{Abbott2005a} and time~\cite{Udem1999a}. Quantum 
metrology exploits fundamental
differences between classical and quantum theories for novel
measurement techniques and enhanced precision~\cite{Lee2002a,Giovannetti2004a}. 
Advantages stem from
several phenomena associated with quantum interferometers, including
nonlocal interference~\cite{Franson1989a,Franson1992a}, phase-insensitive 
interference~\cite{Hong1987a}, phase super-resolution and 
super-sensitivity~\cite{Walther2004a,Mitchell2004a,Resch2007a}, and automatic 
dispersion cancellation~\cite{Franson1992a,Steinberg1992a}.  Unfortunately, 
quantum interferometers require entangled states that are practically 
difficult to create, manipulate, and detect, especially compared to the ease 
of working with robust, intense classical states. 
In the present work, we show that the set of advantages previously associated 
with a quantum interferometer are, in fact, more easily achieved classically. 

\begin{figure}
  \begin{center}
  \includegraphics[width=1.0\linewidth]{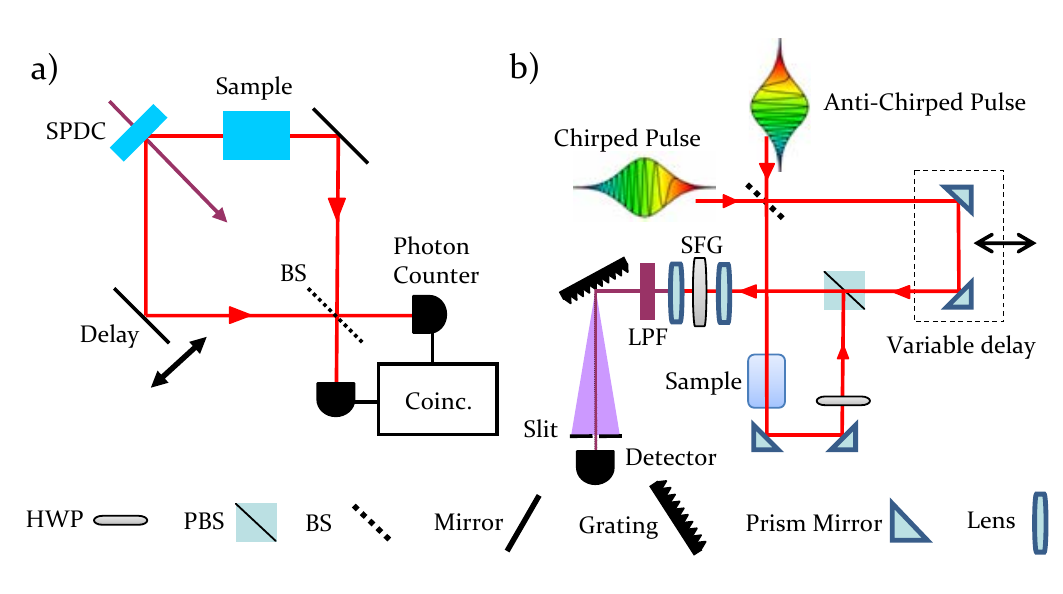}
  \caption{\label{figure1}
    Chirped-pulse interferometry. (a) The HOM interferometer. A
    laser of frequency $2\, \omega_0$creates frequency-entangled photons
    through
    spontaneous parametric down-conversion (SPDC). The photons propagate
    along different paths and are recombined at a beamsplitter (BS) after
    one passes through a sample. The photons can arrive at different
    detectors when the path lengths are unbalanced; when balance is
    achieved, the photons always arrive at the same detector due to
    quantum interference. (b) The chirped-pulse interferometer.
    Oppositely-chirped laser pulses are combined at a beamsplitter (BS).
    The output beams are recombined and focused onto a nonlinear crystal
    after one of the beams has passed through a sample. Type-II sum
    frequency generation (SFG) near the frequency $2\, \omega_0$ as a function 
    of delay is detected using a standard photodiode.}
  \end{center}
\end{figure}

Arguably the best known example of quantum interference was
demonstrated by Hong, Ou, and Mandel~\cite{Hong1987a} (HOM); their 
interferometer is
depicted in fig.~\ref{figure1}a. HOM interference induces strong photon-photon
interactions and is central to optical quantum technologies, including
quantum teleportation~\cite{Bouwmeester1997a} and linear-optical quantum 
computing~\cite{Knill2001a}.
Several characteristics distinguish HOM from classical interference,
such as Michelson’s or Young’s. The HOM signal stems from pairs of
interfering photons and manifests as a dip in the rate of coincident
photon detections which spans the entire coherence length of the
light, as opposed to classical wavelength fringes. It is therefore
inherently robust against path length fluctuations. If the photon
pairs are entangled, the visibility and width of the HOM interferogram
is insensitive to loss~\cite{Steinberg1993a} and 
dispersion~\cite{Steinberg1992a}. Furthermore, HOM
interferometers achieve higher resolution than classical
interferometers using the same bandwidth~\cite{Abouraddy2002a,Nasr2003a}. 
These features are
ideal for precision optical path measurements of dispersive and lossy
materials, implemented by placing the sample in one interferometer arm
and measuring the delay required to restore the dip. A quantum version
of optical coherence tomography~\cite{Fujimoto1995a} (OCT) was proposed and
demonstrated~\cite{Abouraddy2002a,Nasr2003a} to harness these advantages.  

Recently, two proposals~\cite{Erkmen2006a,Banaszek2007a} and one experimental 
demonstration~\cite{Resch2007b} have described classical systems exhibiting 
automatic dispersion cancellation.
Significant drawbacks to these techniques include reliance on
unavailable technology~\cite{Erkmen2006a} or significant 
post-processing~\cite{Banaszek2007a,Resch2007b}. The
experimentally demonstrated technique requires wavelength path
stability; the interference visibility falls precipitously with loss
and is limited to $50\%$ of that possible with the HOM effect.
Alternatively, background-free autocorrelation exhibits enhanced
resolution, phase insensitivity, and robustness against loss, but
notably not automatic dispersion cancellation; this technique has
recently been used in OCT~\cite{Peer2007a}.

In the present work, we describe and
experimentally demonstrate classical interference with \textit{all} of the
metrological advantages of the HOM interferometer. In contrast with
other classical interferometers, the Feynman paths giving rise to the
interference cannot be identified with the spatial paths constituting
the two interferometer arms. Sum-frequency generation (SFG) acts to
directly produce the interference signal from a pair of oppositely
chirped optical pulses with strong classical frequency correlations;
no post-processing or coincidence counting is required. The device can
be understood as a time-reversed HOM interferometer using an
argument~\cite{Resch2007a} based on the corresponding symmetry of quantum 
mechanics.
Remarkably, time reversal converts the quantum interferometer into a
device that can use bright classical laser pulses and achieves a
demonstrated ten-million-fold higher signal. Our general approach
should yield similar improvements in performance when applied to many
other entangled-photon based interferometers.  

Hong-Ou-Mandel-based metrology can be explained using the following 
approach~\cite{Steinberg1992a}. The
wavevector of light in a material can be expanded about a frequency $\omega_0$,
$k(\omega)\approx k(\omega_0)+\alpha\,(\omega-\omega_0)+\beta\,(\omega-\omega_0)^2 +
\ldots$, where $\alpha$ and $\beta$ are material properties describing the
group delay and quadratic group velocity dispersion (GVD),
respectively. Ideal frequency-entangled photon pairs are described by
the state, $\vert\psi\rangle = \int d\Omega\,f(\Omega)\vert\omega_0 + 
\Omega\rangle\vert\omega_0 - \Omega\rangle$, where $f(\Omega)$ is the 
amplitude spectrum. The coincidence rate in the HOM interferometer as a 
function of the relative delay time, $\tau$, is given by~\cite{Steinberg1992a},
\begin{equation}
C(\tau) = \int d\Omega\,\vert f(\Omega)\vert^2\,\left\{
1 - \cos\left[\phi_{rr}(\Omega) - \phi_{tt}(\Omega)\right]\right\}.
\end{equation}
Here $\phi_{rr}(\Omega)$ ($\phi_{tt}(\Omega)$) are the phases associated with 
the amplitude where both photons are reflected (transmitted); the delay
time $\tau=\frac{(L_2 - L_1 + L)}{c}$, where $L_1$ ($L_2$) is the length of 
the sample (delay) arm and $L$ is the length of the sample; and
$\phi_{rr}(\Omega) = L\,(+\alpha \Omega + \beta \Omega^2) - \Omega \tau$
and $\phi_{tt}(\Omega) = L\,(-\alpha \Omega + \beta \Omega^2) + \Omega \tau$,
after removing an irrelevant global phase.  

Since $\phi_{rr}(\Omega)$ and $\phi_{tt}(\Omega)$ have the same dependence 
on the GVD, $\beta$, it is automatically cancelled in the interference signal,
as are all even orders of dispersion. The coincidence rate drops to zero 
for $\tau = \alpha L$, or when the group delay from the material is exactly 
compensated by unequal physical path lengths; this marks the centre of the HOM
dip.  

To understand our chirped-pulse interferometry (CPI) technique, consider
the cross-correlator shown in fig.~\ref{figure1}b as a time-reversed HOM
interferometer (Fig.~\ref{figure1}a). The \textit{detection} of a pair of 
photons with
frequencies $\omega_0 \pm \Omega$ is replaced by the \textit{preparation} 
of a pair of photons with those frequencies; the \textit{preparation} of a 
pump photon of frequency $2\,\omega_0$, which is subsequently down-converted, 
is replaced by the \textit{detection} of a photon of frequency $2\,\omega_0$, 
which had previously been up-converted. The signal in Eq. 1 is built up by 
repeating the experiment with many pairs of photons with frequencies 
distributed according to the spectrum, $\vert f(\Omega)\vert^2$.  

\begin{figure}
  \begin{center}
  \includegraphics[width=1.0\linewidth]{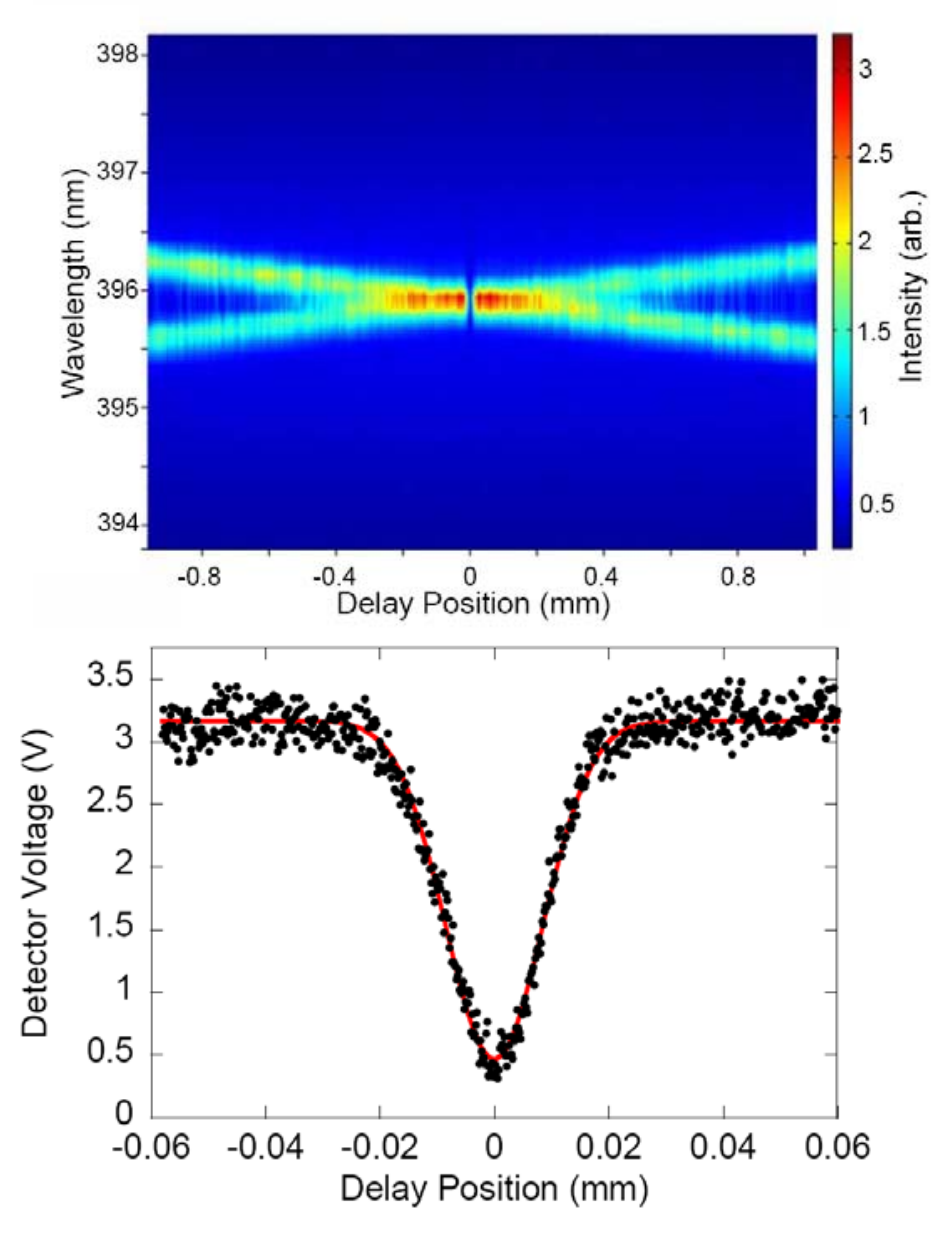}
  \caption{\label{figure2}
    Chirped-pulse interference. (a) Spectrum of the SFG versus
    delay position.  Destructive interference removes the
    cross-correlation signal near zero delay; other features are discussed
    in the text. (b) Using a grating and a slit, we measure the optical
    power at $395.9$nm with a bandwidth of $0.4$nm, as function of delay.
    The signal shows a pronounced dip near zero delay; we use a Gaussian
    fit to measure the visibility $85.2\pm 0.6\%$ and width 
    $19.9\pm 0.2 \mu$m FWHM.}
  \end{center}
\end{figure}

The power of CPI stems from the replacement of photon
pairs by bright classical beams with frequencies $\omega_0 \pm \Omega$. 
The SFG from these beams will contain three distinct frequencies, instead 
of just one: the cross-correlation produces up-converted light at 
$2\,\omega_0$ at a rate proportional to 
$\left\{1 - \cos\left[\phi_{rr}(\Omega)-\phi_{tt}(\Omega)\right]\right\}$ 
(cf. Eq. 1); the autocorrelation produces two new beams at 
frequencies $2\,\omega_0 \pm 2\,\Omega$. A narrow bandpass filter centred 
at $2\,\omega_0$ removes the autocorrelation unless $\Omega$ is small.  

The frequency difference, $\Omega$, is
swept using a pair of oppositely-chirped optical pulses with matched
frequency ramps.  A chirped (anti-chirped) pulse has a frequency that
increases (decreases) linearly in time. We require that the chirp and
anti-chirp are much greater than any dispersion in the interferometer
and stretch the pulses to many times their initial duration.  Under
these conditions, the input frequencies are swept in an
anti-correlated manner such that at any instant only two frequencies,
$\omega_0 \pm \Omega$, are input. (Oppositely-chirped pulses have previously 
been used to
efficiently drive rotational and vibrational transitions in
molecules~\cite{Karczmarek1999a,Xia2003a}). This ramp performs the integration
in Eq. 1
automatically. As an added benefit, chirped pulses can have high peak
intensities yielding correspondingly high frequency-conversion
efficiency. 

We use a modelocked ti:sapphire laser (centre wavelength 790 nm, pulse 
duration 110 fs, average power 2.8 W, repetition rate 80 MHz) as the light
source for the experiment. The polarization of the output is rotated
from vertical to horizontal using a half-wave plate to achieve maximum
diffraction efficiency from our gratings. Our laser light is split
using a 50/50 beamsplitter. Half of the optical power is sent through
a grating-based optical compressor and the other half is sent through
a grating-based optical stretcher~\cite{Treacy1969a,Martinez1988a,
Pessot1987a}. The stretcher applies normal dispersion, creating a chirped 
pulse where the blue lags the red in time, whereas the compressor applies 
anomalous dispersion, creating the anti-chirped pulse where the red lags the 
blue. While the terms stretcher and compressor are commonly used, in our 
experiment both devices stretch our optical pulses. Both stretcher and 
compressor use $30\,\mbox{mm}\times 30\,\mbox{mm}$, $1200$ lines/mm 
gold-coated ruled diffraction gratings, blazed for 800nm.

In the compressor, the gratings are oriented with their faces parallel and 
separated by a distance of $56$cm. The input beam passes over the top of a 
prism mirror; the retro-reflecting mirror is angled slightly downward so that 
the output beam is reflected by the prism mirror. The compressor produces
anti-chirped output pulses $45\pm 0.1$ps long with $9$nm of bandwidth and the 
beam has an average power of $790$mW. 

In the stretcher, the gratings are
oriented with their faces antiparallel and separated by $145$cm. A 1:1
telescope is placed between the gratings and consists of two lenses 
$f\approx 50$cm
separated by $98.5$cm with the first lens placed $9.2$cm after the first
grating. The stretcher produces chirped output pulses $51.2\pm 0.2$ps long 
with $10$nm of bandwidth and the beam has an average power of $870$mW.

Initially, we balanced the stretcher and compressor by sending the
output of the stretcher through the compressor and minimizing the
pulse duration of the output by changing the grating separation in the
compressor. We observed a minimum broadening of $10$\% over pulses
directly from the laser. The differences between the durations of the
chirped and anti-chirped pulses are due to unequal loss of bandwidth
in the stretcher and compressor. They do not reflect different chirp
rates.

The beams of chirped and anti-chirped pulses are injected into the 
cross-correlator as shown in fig.~\ref{figure1}b. To compensate the shorter 
optical path in the compressor as compared to the stretcher, the anti-chirped
pulse arrives at the beamsplitter via a variable delay path (not shown). The 
relevant centre frequency for our experiment is not the centre frequency of 
the pulse, but rather is determined by the temporal overlap of the chirped 
and anti-chirped pulse at the 50/50 beamsplitter. If the chirped pulse
lags (leads) the anti-chirped pulse, the frequency $2\,\omega_0$ is 
red-shifted (blue-shifted) from twice the centre frequency of the laser.
This can be used to make measurements of group delays over a tunable range of
wavelengths, which is difficult to do using HOM interference, since
the entangled photons are typically produced using a fixed frequency
CW laser~\cite{Steinberg1992a,Nasr2003a}. To illustrate this point, we 
combined our pulses such
that the sum of the frequencies corresponded to a wavelength, $395.9$nm, well
separated from half of the centre wavelength of the laser, $395.0$nm. 

The two outputs from the beamsplitter travel different paths through the
cross-correlator. One travels through the delay arm where a
retro-reflector is placed on a motorized translation stage with $40$mm
travel; the other passes through the sample and an achromatic
half-wave plate which rotates the polarization from horizontal to
vertical. The two beams are recombined at a broadband polarizing
beamsplitter cube (PBS).

The output from the PBS is focused by a $5$cm
achromatic lens into a $0.5$mm $\beta$-barium-borate (BBO) optical crystal cut
for collinear type-II degenerate sum-frequency generation. The
sum-frequency beam is then collimated by means of another $5$cm lens.
The infrared light is filtered by means of two dichroic mirrors (not
shown) designed to reflect $395$nm light at $45^\circ$ incidence and to
transmit $790$nm light, as well as a cyan coloured glass low-pass
filter; this is depicted as a low-pass filter (LPF) in fig.~\ref{figure1}b. 

Fig.~\ref{figure2}a shows the measured SFG spectrum as a function of the 
delay. The
cross-correlation signal is clearly observed, but the autocorrelation
signal comprises a broad background barely visible on this scale. For
large delays, the cross-correlation signal contains two
easily-discernable wavelengths spaced symmetrically about $395.9$nm. These
peaks arise from SFG due to the chirped component in the sample arm
and the anti-chirped component in the delay arm, and vice versa. These
different alternatives for producing the cross-correlation signal
constitute the distributed Feynman paths which interfere. The two
wavelengths approach one another as the path length difference
approaches zero, where destructive interference eliminates the
cross-correlation signal.

We filter a bandwidth of $0.4$nm centred at $395.9$nm using a $1200$lines/mm
aluminum-coated diffraction grating followed by a slit. The optical
power is measured using an amplified silicon photodiode. The photodiode 
signal as a function of delay is displayed
in fig.~\ref{figure2}b and clearly shows the interference dip with visibility 
$85.2 \pm 0.6\%$ and FWHM $19.9 \pm 0.6 \mu$m or $133 \pm 1$fs.

All visibilities were calculated without background
subtraction; however, our photodiode registered a bias ranging from
$−30$mV to $−40$mV when in the dark. We measured this bias for every data
set and subtracted the negative value from our measured voltage. Note
that this bias correction lowers our reported visibilities.  

\begin{figure}
  \begin{center}
  \includegraphics[width=1\linewidth]{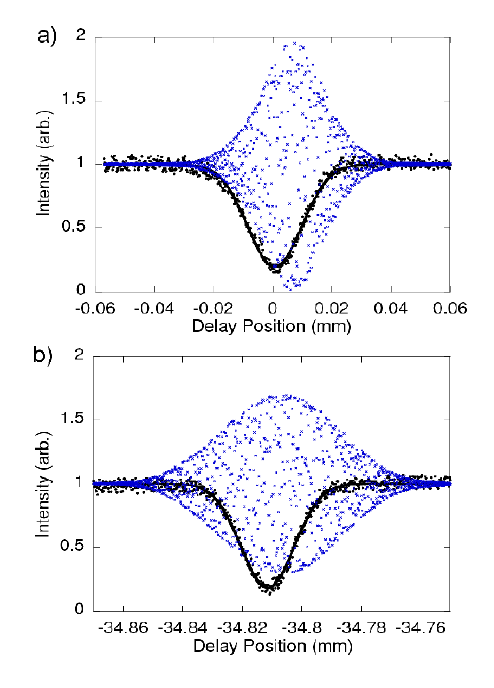}
  \caption{\label{figure3}
    Automatic dispersion cancellation in chirped-pulse
    interferometry. These data show a direct comparison of the
    chirped-pulse interference signal (black circles) and standard
    white-light interference using the chirped pulse (blue x’s) when (a) no
    additional glass and (b) $80.60\pm 0.05$mm of calcite (oriented for 
    o-polarization) and $28.93\pm 0.04$mm of BK7 glass are placed in the 
    sample arm. Note that the small
    offset between the CPI and white-light interference is due to the
    birefringence of the up-conversion crystal.
    }
  \end{center}
\end{figure}

Our measured visibility easily surpasses the $50\%$
limit commonly attributed to any classical analogue of HOM
interference. This classical limit applies only to the visibility of
the coincidence rate (or correlation) between two square-law
photodetector signals showing no individual 
interference~\cite{Hong1987a,Paul1986a}. Although
both SFG and coincidence detection measure correlations, the SFG
signal depends on the product of the electric fields, as opposed to
intensities; thus our detection scheme avoids this constraint. In
practice, the background from the autocorrelation does limit the
visibility, but it can be arbitrarily close to $100\%$ with large chirp
and narrow filtering. Alternatively, one could achieve $100\%$ visibility
by removing the small band of frequencies responsible for the
autocorrelation background from the chirped and anti-chirped pulses;
this has the drawback of distorting the interferogram.  

The optical power corresponding to $1$V on our detector was measured to be 
$1.5 \mu$W at $395$nm, thus our measured signal of $4.5 \mu$W corresponds to 
about $10^{13}$ photons/s. The highest reported coincidence 
rate~\cite{Altepeter2005a} from a photon pair source is $2\times 10^6$ Hz 
while the rate is typically orders of magnitude lower in a HOM interferometer. 
Our signal is a demonstrated 7 orders of magnitude higher than what can be 
achieved in a HOM interferometer using state-of-the-art photon-pair sources.

To demonstrate automatic
dispersion cancellation, we took two data sets: one with significant
dispersive material in the sample arm, $80.60 \pm 0.05$mm of calcite and 
$28.93\pm 0.04$mm of BK7 glass, and one without. (The dispersive properties 
of calcite and BK7 do not cancel, rather their effects are cumulative). In 
each configuration we measured chirped-pulse and white-light
interferograms. To observe white-light interference, we sent the
chirped pulse through the interferometer, placed a polarizer at $45^\circ$
before the nonlinear crystal, and directly detected the transmitted
infrared light. The resulting interferograms are shown in 
fig.~\ref{figure3}a and \ref{figure3}b. The CPI widths and centres were 
obtained by a Gaussian fit whereas the white-light interference 
characteristics were obtained via the Hilbert transform 
method~\cite{Fercher2003a}.  

With no sample, fig.~\ref{figure3}a, we observe $143 \pm 2$fs FWHM for the
chirped-pulse dip and $173 \pm 1$fs FWHM for the white-light interference
pattern.  By comparing the widths we see that the chirped-pulse signal
has $17\%$ better resolution. Theory predicts an increase in resolution
of $29\%$ (assuming Gaussian bandwidths); we attribute the difference to
the acceptance bandwidth of our SFG crystal, the offset of our
chirped-pulse average wavelength from our pulse centre wavelength, and
the slightly unequal bandwidths of our chirped and anti-chirped
pulses.  With the dispersive elements, fig.~\ref{figure3}b, we observed 
$140 \pm 2$fs
FWHM for chirped-pulse interference and $303 \pm 2$fs FWHM for white-light
interference. Dispersion clearly increased the width of the
white-light interference pattern by $75\%$; the width of the
chirped-pulse interference pattern remained essentially unchanged due
to dispersion cancellation.

To show that CPI accurately determines group delays, we measured shifts in the 
centre of the interference of $34811.9\pm 0.3 \mu$m and $34813.80\pm 0.3\mu$m 
for the chirped-pulse dip and white-light fringes respectively. These agree 
well with theoretical shifts of $34816\pm 20 \mu$m and $34822\pm 20 \mu$m,
calculated from the group delays at $791.8$nm and $790$nm, respectively.
Uncertainties in the theory result from errors in the measurement of
sample thickness.  

\begin{figure}
  \begin{center}
  \includegraphics[width=0.8\linewidth]{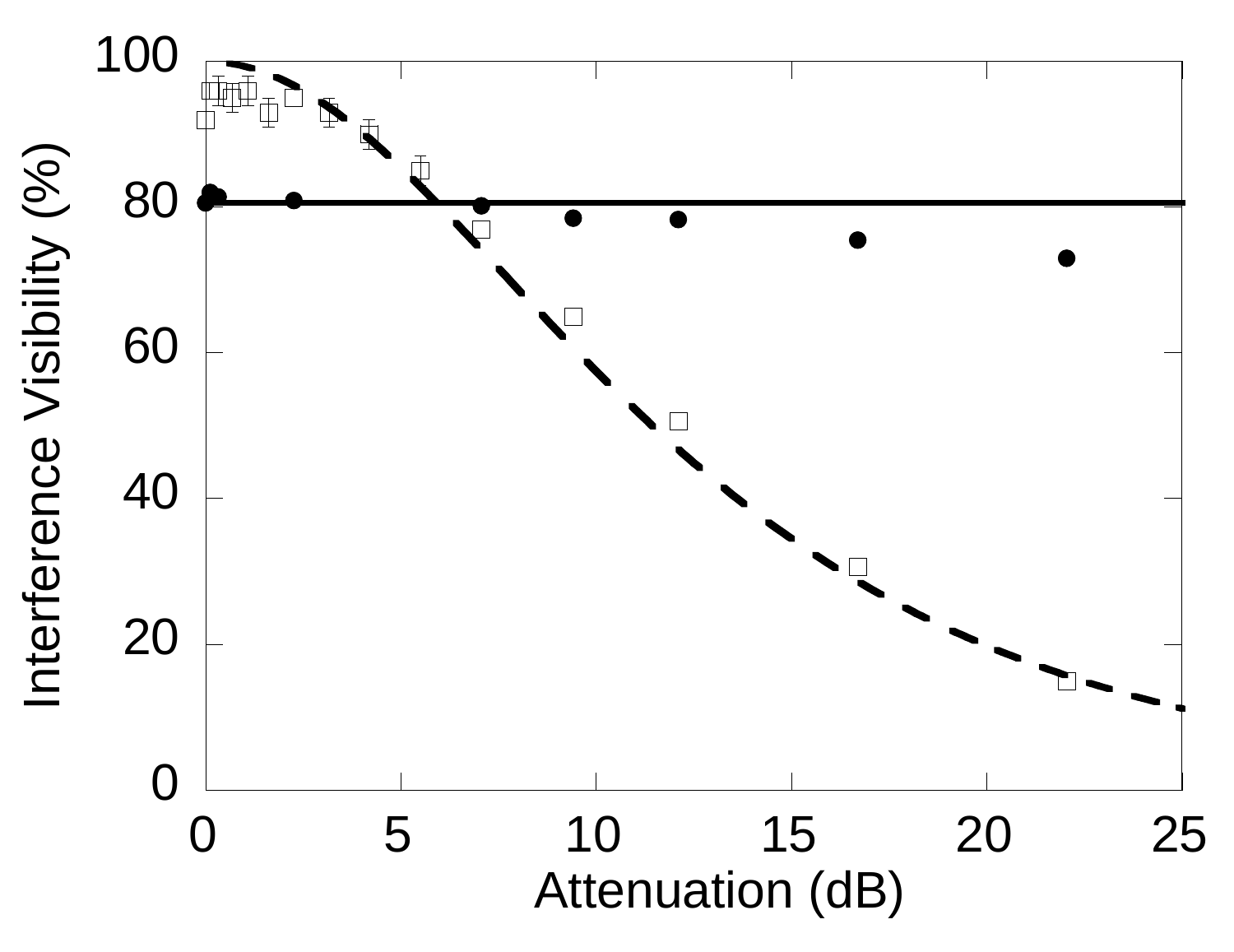}
  \caption{\label{figure4}
    Visibility versus unbalanced loss. These data show the
    visibility of the chirped-pulse interference dip (closed circles,
    solid line theory) and white-light interference (open squares, dashed
    line theory) as a function of loss in the sample arm introduced by
    rotating the half-wave plate (Fig.~\ref{figure1}b). The chirped-pulse 
    interference
    visibility shows far more resilience to loss. Error bars indicate
    statistical errors of $1$ s.d.}
  \end{center}
\end{figure}

A further advantage of CPI as compared to
white-light interferometry is the insensitivity of the visibility to
unbalanced loss in the interferometer arms; loss will, however, reduce
the overall output intensity, and thus the signal, in both cases. We
measured both visibilities as a function of attenuation in the sample
path. Rotating a half-wave plate in the sample path enables continuous
adjustment of the loss at the polarizing beam splitter. The results of
these measurements are shown in fig.~\ref{figure4}. The visibility in the
chirped-pulse interference is far more robust than the white-light
interference, dropping only slightly at high attenuation due solely to
background. This insensitivity can be explained by noting that in CPI
the loss is common to both interfering Feynman paths even though it is
localized in one physical path.

Chirped-pulse interferometry features
all of the metrological advantages of Hong-Ou-Mandel interference with
vastly higher signal levels. CPI achieves this without the inherent
disadvantages of entangled photon sources and single-photon detection.
Increasing the laser bandwidth and the spectral acceptance of
sum-frequency generation~\cite{Carrasco2006a} will make CPI resolution 
competitive with that in optical coherence tomography~\cite{Fujimoto1995a,
Drexler2004a}. Automatic dispersion
cancellation, enhanced resolution, and insensitivity to loss and path
length fluctuations promise to make CPI a superior imaging technology,
especially for dispersive and lossy media, e.g. biological specimens
and photonic devices. More generelly, our work emphasizes the
importance of delineating truly quantum effects from those with
classical analogue~\cite{Steinberg1992a,Bennink2002a,Ferri2005a}, and shows 
how quantum insights can inspire novel classical technologies.  Our approach 
provides an avenue into previously untapped potential of classical 
interferometry.

We thank D. Strickland for sharing invaluable
expertise on pulse compression techniques and K. Bizheva and G. Weihs
for important discussions and loaning equipment. We thank J.
Sanderson, J. Lundeen, M. Mitchell, J. Gambetta, A. White, and A.
Steinberg for helpful comments. This work was supported by NSERC, and
CFI. D.B. and R.K. acknowledge financial support from the Mike and
Ophelia Lazaridis Fellowship and IQC, respectively.  

\bibliography{cpi}

\end{document}